# Distinguishing Mechanisms Underlying EMT Tristability

Dongya Jia[1,2], Mohit Kumar Jolly[1,3], Satyendra C. Tripathi[7], Petra Den Hollander[8], Bin Huang[1,6], Mingyang Lu[1], Muge Celiktas[7], Esmeralda Ramirez-Peña[8], Eshel Ben-Jacob[1,12*], José N. Onuchic[1,4,5,6], Samir M. Hanash[7,9], Sendurai A. Mani[10,11], Herbert Levine[1,2,3,4,5]

[1]Center for Theoretical Biological Physics, [2]Systems, Synthetic and Physical Biology Program, Department of [3]Bioengineering, [4]Biosciences, [5]Physics and Astronomy and [6]Chemistry, Rice University, Houston, TX 77005, USA

[7]Department of Clinical Cancer Prevention, University of Texas MD Anderson Cancer Center, Houston, TX 77030, USA

[8]Department of Translational Molecular Pathology, University of Texas MD Anderson Cancer Center, Houston, TX 77030, USA

[9]Red and Charline McCombs Institute for the Early Detection and Treatment of Cancer, University of Texas MD Anderson Cancer Center, Houston, TX 77030, USA

[10]Department of Translational Molecular Pathology, University of Texas MD Anderson Cancer Center, Houston, TX 77030, USA

[11]Metastasis Research Center, University of Texas MD Anderson Cancer Center, Houston, TX 77025, USA

[12]School of Physics and Astronomy, Tel-Aviv University, Tel-Aviv 69978, Israel

* Deceased on June 5, 2015

Correspondence to:

Herbert Levine, e-mail: herbert.levine@rice.edu

**Abstract**

Background: The Epithelial-Mesenchymal Transition (EMT) endows epithelial-looking cells with enhanced migratory ability during embryonic development and tissue repair. EMT can also be co-opted by cancer cells to acquire metastatic potential and drug-resistance. Recent research has argued that epithelial (E) cells can undergo either a partial EMT to attain a hybrid epithelial/mesenchymal (E/M) phenotype that typically displays collective migration, or a complete EMT to adopt a mesenchymal (M) phenotype that shows individual migration. The core EMT regulatory network - miR-34/SNAIL/miR-200/ZEB1 - has been identified by various studies, but how this network regulates the transitions among the E, E/M, and M phenotypes remains controversial. Two major mathematical models – ternary chimera switch (TCS) and cascading bistable switches (CBS) - that both focus on the miR-34/SNAIL/miR-200/ZEB1 network, have been proposed to elucidate the EMT dynamics, but a detailed analysis of how well either or both of these two models can capture recent experimental observations about EMT dynamics remains to be done.

Results: Here, via an integrated experimental and theoretical approach, we first show that both these two models can be used to understand the two-step transition of EMT - E→E/M→M, the different responses of SNAIL and ZEB1 to exogenous TGF-β and the irreversibility of complete EMT. Next, we present new experimental results that tend to discriminate between these two models. We show that ZEB1 is present at intermediate levels in the hybrid E/M H1975 cells, and that in HMLE cells, overexpression of SNAIL is not sufficient to initiate EMT in the absence of ZEB1 and FOXC2.

Conclusions: These experimental results argue in favor of the TCS model proposing that miR-200/ZEB1 behaves as a three-way decision-making switch enabling transitions among the E, hybrid E/M and M phenotypes.

**Keywords**

Epithelial-Mesenchymal Transition, EMT, Ternary chimera switch, TCS, Cascading bistable switches, CBS, ZEB1, FOXC2

**Background**

The Epithelial-to-Mesenchymal Transition (EMT) and its reverse process - Mesenchymal-to-Epithelial Transition (MET) - play critical roles during embryonic development and tissue repair. This process can also be utilized by cancer cells to acquire properties similar to stem cells, to become drug-resistant and to obtain enhanced migratory abilities (1,2). Epithelial cells from a primary tumor can undergo EMT to lose cell-cell adhesion and acquire mesenchymal invasive properties (3). These transitioned cells can enter blood vessels and migrate as Circulating Tumor Cells (CTCs) (4). Eventually, the CTCs may exit the vasculature at a distant organ, undergo MET, seed and thereby form a secondary tumor or metastases (1), which are the cause of 90% of cancer-related deaths (5). Hence decoding the operating principles of EMT is crucial to unveil the mechanism of enhanced metastasis and therapeutic failure.

Emerging evidence shows that in addition to epithelial (E) and mesenchymal (M) phenotypes, cells can acquire a hybrid epithelial/mesenchymal (E/M) phenotype (also referred to as 'partial EMT') that has combined traits of epithelial (cell-cell adhesion) and mesenchymal (invasion) phenotypes (6–11). Consequently, cells in the hybrid E/M phenotype can migrate collectively as a cluster (12). These clusters, which originate from the primary tumor front, can display up to 50 times higher tumor-formation potential as compared to individually migratory mesenchymal cells (13). In addition, the hybrid E/M phenotype has been suggested to be more correlated with stem-like properties (14,15) and chemoresistance (16) as compared to epithelial and mesenchymal phenotypes. Thus characterizing the hybrid E/M phenotype can contribute to a full understanding of the role of EMT in metastasis and chemoresistance.

The signaling network orchestrating EMT is complex. For example, EMT can be triggered by many signaling pathways such as TGF-β, Notch and Wnt (17), and different mechanical factors such as extracellular matrix density (18) and mechanical stress (19). On the other hand, EMT can be repressed by tumor suppressors such as p53 (20), transcription factors such as GRHL2 (21) and OVOL (22), and microRNA families such as miR-200 and miR-34 (23). Despite the complexity of the signaling network, there appears to exist a 'hub' that functions as the master regulator of EMT. This 'hub' consists of two interconnected mutually inhibitory feedback loops between microRNAs and transcription factors – one between the miR-34 family and SNAIL1/2, and the other between the miR-200 family and ZEB1/2 (24–26). High levels of miR-200 and miR-34 are associated with an epithelial phenotype, and high levels of ZEB1/2 and SNAIL are associated with a mesenchymal phenotype (27–30).

Recently, two different mathematical models - the ternary chimera switch (TCS) (31) and the cascading bistable switches (CBS) (32) – have been proposed to elucidate phenotypic transitions during EMT (**Figure 1A, B,** see **SI Section 1&2** for the details of model formulation). Both models focus on the aforementioned EMT regulatory circuits – miR-34/SNAIL and miR-200/ZEB (which will be referred to as miR-200/ZEB1 hereafter since the experimental results discussed later focus on ZEB1), and can explain the two-step transitions during EMT - E→E/M→M (**Figure 1C, D**) (31,32). Despite these similarities, however, the TCS and CBS models differ on an important aspect – the role of ZEB1 during EMT. The TCS model proposes that ZEB1 levels are trimodally distributed among E, E/M and M phenotypes, and intermediate levels of ZEB1 are required to maintain the hybrid E/M phenotype (31). In other words, an upregulation of ZEB1 levels is required to both initiate partial EMT (E→E/M) and complete EMT (E/M→M) (**Figure 1E**). In contrast, the CBS model proposes that ZEB1 levels are bimodally distributed, such that the levels of ZEB1 in E and hybrid E/M phenotypes are relatively low and not significantly different, while that in M phenotype are relatively high (32) (**Figure 1F**). These differences in the proposed role of ZEB1 exist between TCS and CBS models because of the different characterizations of the EMT regulatory circuits by each model. The TCS model proposes the miR-200/ZEB1 circuit to be the three-way decision-making switch of EMT and the miR-34/SNAIL to be a monostable noise-buffering integrator (31). In contrast, the CBS model proposes that both the miR-34/SNAIL circuit and the miR-200/ZEB1 circuit are bistable switches, with miR-34/SNAIL responsible for the switch from E to E/M and miR-200/ZEB1 responsible for the switch from E/M to M (32).

Despite modeling the same circuit, a key reason why these two models have different predictions is that the TCS model, but not the CBS model, includes the self-activation of ZEB1. It is in fact well-known that including self-activation in toggle switch systems can lead to tristability (33–35). In this study, we first confirm the importance of the ZEB1 self-activation based on recently published work on the self-enforcing CD44s/ZEB1 feedback loop. The trimodal distribution of ZEB1 from the TCS modeling analysis is further supported by gene expression data from NCI-60 cell lines and the immunofluorescence (IF) results of the hybrid E/M non-small cell lung cancer (NSCLC) cell line – H1975.

Next, we show that both the TCS model and the CBS model can recapitulate currently published experimental results such as (a) EMT is a two-step transition – E to hybrid E/M to M, (b) a lower TGF-β concentration is required for the activation of SNAIL compared to that for ZEB1, (c) the transition from E to E/M is reversible while the transition from E/M to M is largely irreversible in MCF10A cells. Hence, much of the existing data does not distinguish between these two possibilities.

Hence, we present novel experimental data that is capable of discriminating between the two models. We show that overexpression of SNAIL in HMLE cells cannot induce EMT without the expression of ZEB1 (and FOXC2, which is another important mediator of EMT and is known to regulate ZEB1 (36)). This observation suggests that the miR-34/SNAIL circuit may not be responsible for the switch from E to E/M phenotypes and is therefore more congruent with predictions from the TCS model.

**Results**

**The CD44s/ZEB1 feedback loop enables the miR-200/ZEB1 circuit to function as a three-way switch.**

The TCS model incorporates a direct ZEB1 self-activation link based on the contribution of ZEB1 in stabilizing the SMAD complexes; this link can reinforce ZEB1 levels that are raised via the TGF-β pathway (25). Here, we include in the model recently published evidence regarding a self-reinforcing feedback loop of ZEB1 via ESRP1 and CD44s (37) (**Figure 2A**), and show that this feedback loop indeed enables the tristability of the miR-200/ZEB1 circuit (see **SI Section 3** for the model formulation of the CD44s/ZEB1 feedback loop and **Table S1** for relevant parameters).

CD44, a crucial stem cell marker, has two major isoforms – CD44s (the CD44 standard isoform) and CD44v (CD44 variant isoform) (38). Whereas the total amount of CD44 is maintained largely unchanged during EMT, the isoform switch from CD44v to CD44s is essential for the progression of EMT (39,40), and CD44s can upregulate ZEB1 (37). This isoform switch is regulated by a splicing factor - epithelial splicing regulatory protein 1 (ESRP1) - which promotes the splicing of CD44v and inhibits that of CD44s (39). The transcription of ESRP1 can be directly inhibited by ZEB1, and therefore ZEB1 can upregulate itself by promoting the production of CD44s (37,39).

To analyze the effect of the CD44s/ZEB1 feedback loop on the behavior of the miR-200/ZEB1 circuit, we calculated a bifurcation diagram (**Figure 2A**) to illustrate the existence of and transitions among the different stable states. The CD44s/ZEB1 feedback loop enables the miR-200/ZEB1 circuit to acquire three stable states (phenotypes) – (low ZEB1, high ESRP1) corresponding to the E phenotype, (medium ZEB1, medium ESRP1) corresponding to the hybrid E/M phenotype and (high ZEB1, low ESRP1) corresponding to the M phenotype. The hybrid E/M phenotype is not seen upon removal of the CD44s/ZEB1 feedback loop (**Figure S1**), demonstrating that self-activation in a toggle switch is critical for attaining more than two stable states (33–35). To understand the diverse EMT-inducing results (e.g., by hypoxia, which can upregulate both SNAIL and ZEB1 or by TGF-$\beta$, which mainly upregulates SNAIL), we calculated the phase diagram where the miR-200/ZEB1 circuit is driven by two independent signals – a transcriptional inhibition signal $S_1$ on miR-200 and a transcriptional activation signal $S_2$ on ZEB1 (**Figure 2B**). Different combinations of $S_1$ and $S_2$ allow 7 possible phases – 3 monostable phases - {E}, {E/M} and {M}, 3 bistable phases - {E/M, E}, {E/M, M} {E, M} and 1 tristable phase {E, E/M, M} (**Figure 2B**). The bistable and tristable phases imply the co-existence of multiple stable states. For example, the tristable phase {E, E/M, M} implies that all three phenotypes – E, E/M and M can exist in the given range of $S_1$ and $S_2$, i.e. cells can switch back and forth among these three phenotypes. Depending on the details of EMT induction, cells tend to follow different trajectories in the phase diagram and thereby undergo different phenotypic transitions. The tristability of the miR-200/ZEB1 circuit enabled by the CD44s/ZEB1 feedback loop is quite robust to parameter perturbation (**SI Section 4, Figure S2**).

Next, to investigate the gene expression levels of ZEB1 across E, hybrid E/M and M cells, we analyze the NCI-60 panel of cell lines that have been classified into E, E/M, and M phenotypes on a population level based on ensemble CDH1/VIM ratio (29). CDH1 encodes the epithelial marker E-cadherin, loss of which is a hallmark of EMT. VIM encodes vimentin, which is a commonly used marker for the mesenchymal phenotype. Both ZEB1 and ESRP1 show three different expression levels across the E, E/M and M cell lines, and the difference among these levels is statistically significant (**Figure 2C**). These results are consistent with the prediction from the TCS model, which proposes that ZEB1 (**Figure 2A, B)** and ESRP1 levels (**Figure S3**) in the hybrid E/M phenotype are different from that of the E phenotype. Further, across the entire NCI-60 panel, ESRP1 expression negatively correlates with ZEB1 and VIM, and positively correlates with CDH1 (41–43). These observations, together with the reported strong negative correlation between the expression of ZEB1 and ESRP1 in the lung, breast and pancreatic cancer patient samples (37), suggest that the ZEB1-ESRP1 axis is active across multiple cancer types (**Figure 2C**).

In addition to the expression analysis of ZEB1 from the NCI-60 cell line panel, we further conducted experiments in three NSCLC cell lines – H820, H1975 and H1299 - that have been characterized as epithelial, hybrid E/M and mesenchymal respectively (44). Notably, the hybrid E/M H1975 cells stain for both E-cadherin and vimentin at a single-cell level and display collective migration (44,45). The immunofluorescence staining of the H1975 cells clearly shows the expression of ZEB1 in the nucleus concomitantly with E-cadherin on the membrane, which demonstrates that ZEB1 is present in the hybrid E/M phenotype as compared to the lack of ZEB1 in the epithelial H820 cells (**Figure 2D, Figure S4**). The intermediate levels of ZEB1 in hybrid H1975 cells are further verified by RT-PCR (**Figure 2E**) and Western blot (**Figure 2F, Figure S5**). Notably, the difference of SNAIL levels (3 fold) is much smaller compared with the difference of ZEB1 (20 fold) between epithelial cells - H820 and H1437 and hybrid E/M cells – H1975 (**Figure 2E, F**), suggesting ZEB1 levels may correspond strongly with maintaining these phenotypes, at least in these cell lines. Interestingly, the levels of SNAIL in mesenchymal cells – H1299 and H2030 were lower than those in epithelial cells - H820 and H1437 (**Figure 2E, F**), further arguing for a more critical role of ZEB1 in maintaining a mesenchymal phenotype, potentially via multiple feedback loops (37,46).

These results indicate that ZEB1 self-activation (direct or indirect) should be considered an integral part of the core EMT circuit, thereby implying that TCS model captures the biological mechanisms more precisely than the CBS one. The parameter region of the bifurcation and phase diagram (**Figure 2A, B**) here is quite consistent with that for the miR-200/ZEB1 circuit with direct ZEB1 self-activation (see Fig. 4 in (31) for comparison). For simplicity, the CD44s/ZEB1 feedback loop is represented by a direct ZEB1 self-activation in the following analysis.

Next, we focus on published experimental data that has been claimed to support the CBS model over the TCS model. These experimental results are – (a) during TGF-β treatment of MCF10A cells, the concentration of exogenous TGF-β required to increase the SNAIL abundance is lower than that needed for a comparable increase in ZEB1 abundance (32), and (b) MCF10A cells that attain a hybrid E/M phenotype revert to being epithelial upon removal of exogenous TGF-β, but cells that attain a mesenchymal phenotype fail to do so (32). In the following two sections, we show that both the CBS and the TCS models can recapitulate these two experimental observations.

**Different responses of SNAIL and ZEB1 to the EMT-inducing signal, exogenous TGF-$\beta$**

To test whether the TCS model can recapitulate the different TGF-β-dose responses of SNAIL and ZEB1, we apply an EMT-inducing signal $S_A$ on SNAIL to mimic the induction of exogenous TGF-β, and analyze the steady-state responses of SNAIL and ZEB1 to different levels of $S_A$. To investigate how differently SNAIL and ZEB1 respond to TGF-β induction at a single-cell level compared with that at a cell population level, we design two different kinds of stochastic simulations. First, at a single-cell level, in absence of a detailed understanding of the gene expression noise in regulating EMT, we consider the effects of white noise. Secondly, to investigate population-averaged results, we include cell-cell variability, which is represented by randomly assigned parameters to each cell.

The TCS model results show that the levels of TGF-β ($S_A$) required for the activation of SNAIL ($S_A = 20 * 10^3$ molecules) is indeed lower as compared to that for ZEB1 ($S_A = 40 * 10^3$ molecules) on both the single-cell (**Figure 3A-C**) and population levels (**Figure 3D-F**). The difference in the upregulation of SNAIL and ZEB1 at different levels of TGF-β might be attributed to the different numbers of binding sites on 3'UTR of mRNA for the corresponding microRNAs – 6 or more binding sites of miR-200 family on ZEB1 mRNA and 2 binding sites of miR-34 family on SNAIL mRNA (47,48). The different endogenous levels of miR-34 and miR-200 family members may also affect the difference in ZEB1 and SNAIL levels. In addition, tristability is more apparent for the ZEB1 mRNA levels than for the ZEB1 protein levels. Again, this can be attributed to the microRNA-mediated repression. Therefore, with intermediate levels of ZEB1 mRNA and miR-200 present in the hybrid E/M phenotype, the ZEB1 protein does not necessarily reach the intermediate level, because of the strong repression by miR-200; this results in the relatively less separated E and E/M phenotypes for ZEB1 protein levels. The fraction of each group with the ZEB1 mRNA levels - low, medium, high - can be adjusted by the levels of TGF-β ($S_A$) (**Figure S6**).

Next, we compare the responses of ZEB1 on the single-cell and population-averaged levels; we find that ZEB1 mRNA levels display a more significant trimodal distribution in the single-cell simulation compared with that from the population-averaged result (**Figure 3C**). The cell-to-cell variability, as reflected by parameter randomization, tends to smoothen the distributions of ZEB1 mRNA in the E and the hybrid E/M phenotypes, potentially because the difference of ZEB1 mRNA levels between E and E/M is smaller as compared to that between E/M and M, as reflected by the simulations of single cells undergoing EMT (**Figure 3A**). The difference of ZEB1 mRNA levels in E, E/M and M phenotypes shown here should not be compared with the gene expression data from NCI-60 (**Figure 2C**), which is not time-course data for individual cells undergoing EMT. Since the existing experimental data on population measurements only yield average values, this may help explain why ZEB1 mRNA appears to be bimodally distributed (32). Consequently, the averaged ZEB1 levels may not be able to argue strongly in favor of one mathematical model over the other.

**The inhibition of miR-34 by ZEB1 as well as autocrine TGF-β signaling stabilizes the mesenchymal phenotype**

To decode possible mechanisms of the irreversible nature of the transition from hybrid E/M to M, we focus on the feedback loop from the miR-200/ZEB1 circuit to the miR-34/SNAIL circuit – the inhibition of miR-34 by ZEB1 (**Figure 4A**). To understand the effect of this link on the transitions among E, E/M and M, we calculate the bifurcation diagram of the complete circuit – miR-34/SNAIL/miR-200/ZEB1 - driven by EMT-inducing signal $S_A$ on SNAIL for varying strengths of the inhibition of miR-34 by ZEB1 (**Figure 4B**). The stronger the inhibition of miR-34 by ZEB1, the smaller the level of EMT-inducing signal $S_A$ required for the cells to transition into and maintain a M phenotype (**Figure 4B**). In addition, a stronger inhibition of miR-34 by ZEB1 can decrease the duration of the hybrid E/M phenotype and can therefore promote a quicker transition from the hybrid E/M phenotype to the M phenotype during temporal dynamic simulation (**Figure S7**).

To characterize the relative stability of E, E/M and M states, we calculate the effective landscape of EMT at different strengths of the inhibition of miR-34 by ZEB1. Here, the external activation signal $S_A$ is chosen at $50*10^3$ molecules, which enables the existence of all three stable states – E, E/M and M in all cases with different strengths of the inhibition of miR-34 by ZEB1 (**Figure 4B**). When there is no feedback from ZEB1 to miR-34, cells are mainly bimodally distributed in either the E or the E/M phenotype (left panel in **Figure 4B**). An intermediate strength of the inhibition of miR-34 by ZEB1 enables all three phenotypes to be attained (middle panel in **Figure 4B**). When the strength of the inhibition of miR-34 by ZEB1 is further increased, the M phenotype becomes the dominant one, thus cells in this case can maintain the mesenchymal phenotype without reverting (right panel in **Figure 4B**). Therefore, a strong inhibitory feedback from ZEB1 to miR-34 stabilizes the mesenchymal phenotype and contributes to the irreversible transition from the hybrid E/M to the M phenotype.

Another possible mechanism accounting for the 'irreversibility' of the mesenchymal phenotype can be the autocrine TGF-β/miR-200 loop, as suggested by the CBS model. Here, we evaluate the effect of the TGF-β/miR-200 loop based on the TCS framework (**Figure 5A**) and show that removal of exogenous TGF-β cannot induce MET but instead the circuit is able to maintain the mesenchymal phenotype (Phase {M} in **Figure 5B**) as long as the innate production rate of endogenous TGF-β is high. This bifurcation diagrams (**Figure 5C, D**) indicate that the 'irreversibility' of mesenchymal phenotype to switch back to a hybrid E/M phenotype depends directly on the production rate of endogenous TGF-β.

In summary, both the TCS and the CBS models can explain currently published experimental results regarding EMT – (a) EMT is a two-step process, from E to E/M to M. (b) The concentration of exogenous TGF-β required for the activation of SNAIL is lower than that for ZEB1, (c) The mesenchymal phenotype can be maintained without reverting when exogenous TGF-β is removed due to certain levels of endogenous TGF-β. In addition, we show that the inhibition of miR-34 by ZEB1 can be responsible for the irreversibility of a complete EMT. These EMT mathematical models are helpful to the extent that they can continue to explain and predict the regulatory effects of newly identified EMT players. To this end, we focus on adding recently identified EMT regulator – FOXC2, which plays a crucial role in EMT progression, to both the TCS and the CBS frameworks and test whether these two models can elucidate the function of FOXC2 in regulating EMT.

**Overexpression of SNAIL is insufficient to initiate EMT in absence of ZEB1 and FOXC2**

The transcription factor FOXC2 serves as a key mediator in regulating EMT and linking EMT with stem-like properties and with metastatic competence (36,49–51). FOXC2 expression can be upregulated by multiple EMT-inducing signals, such as TGF-β1, Snail Goosecoid and Twist. It has been shown that FOXC2 expression is required to maintain the mesenchymal phenotype, the invasive properties and the stem cell-enrichment of HMLE cells following EMT induction (36,49).

FOXC2 regulates EMT through its interactions with core EMT components – SNAIL and ZEB1. Overexpression of SNAIL significantly upregulates the expression of FOXC2 while overexpression of FOXC2 does not affect SNAIL levels (49,52) (**Figure S8**). This suggests that SNAIL functions as an upstream regulator of FOXC2. In addition, FOXC2 directly upregulates the expression of ZEB1 by binding to its promoter region (51). Here, we have expanded both the TCS and CBS models to include transcriptional regulation by FOXC2 (**Figure 6A, Figure S9A,** see **SI Section 5&6** for model formulation and **Table S2&3** for parameters).

To analyze the EMT-inducing behaviors of the TCS model in the presence and absence of FOXC2, we study the effect of two external signals - $S_A$ representing an EMT-inducing signal on SNAIL, and $S_I$ representing an inhibitory signal on FOXC2. The presence of FOXC2 ($S_I = 0$) enables tristability of EMT and accounts for the two-step transition - from E to E/M and from E/M to M (**Figure 6B**, top panel). The absence of FOXC2 ($S_I = 2 * 10^5$ molecules) results in low levels of ZEB1 mRNA and consequently the maintenance of epithelial phenotype (**Figure 6B**, bottom panel) irrespective of the high levels of SNAIL (**Figure S10**).

To test the above-mentioned predictions from the TCS model, we examined the immortalized HMLE cells to assess the role of FOXC2 knockdown (FOXC2-KD) during EMT. Here we measure the protein levels of canonical EMT markers in the HMLE-Snail cells, that have already undergone a complete EMT via SNAIL overexpression, in the presence and absence of FOXC2. We find that FOXC2-KD in HMLE-Snail cells eliminates the expression of ZEB1, vimentin and fibronectin while it restores the expression of E-cadherin, thus inducing a complete MET irrespective of SNAIL overexpression (**Figure 6C**). This experimental observation is consistent with the prediction from the TCS model, but not with the prediction from the CBS model that an EMT-inducing signal can still drive the transition from an epithelial state to a hybrid E/M state upon FOXC2-KD (**Figure S9**).

Is it possible that FOXC2 acts together with SNAIL and together can induce (partial) EMT even without ZEB1? This would argue against TCS and would instead be consistent with a modified version of CBS. We do not think this is supported by existing data. Observations in LNCaP and DU145 cells show that ZEB1 mediates the effect of FOXC2 on tumor-initiating potential and drug resistance (50), traits that are often correlated with EMT (14,53–56). Decreased expression of ZEB1 in PANC-1 cells, which express both epithelial and mesenchymal markers (57) and are thus likely to be hybrid E/M cells, results in a complete MET despite upregulation of SLUG and SNAIL in TGF-β treated cells (58). Moreover, knockdown of ZEB1, but not necessarily of SNAIL and SLUG, had pronounced effects in cells losing their EMT-like properties (59). Therefore, we believe that the most consistent interpretation of the data is that SNAIL (even with FOXC2) is insufficient without ZEB1 and conversely both ZEB1 and FOXC2 are needed (**Figure 6D**) and contribute to different aspects of driving EMT such as repression of epithelial program (ZEB1 is a transcriptional repressor (60)) and activation of mesenchymal program (FOXC2 is a transcriptional activator (51)). However, further experiments such as overexpression of ZEB1 in FOXC2 knockdown and overexpression of FOXC2 in ZEB1 knockdown cells will be crucial in further delineating the mechanisms of EMT dynamics and distinguishing the principles underlying EMT tristability.

**Discussion**

Phenotypic transitions among epithelial, hybrid E/M, and mesenchymal phenotypes endow cancer cells with rich plasticity to metastasize and form secondary tumors. To elucidate the operating principles of EMT/MET, two conceptual frameworks - Ternary Chimera Switch (TCS) and Cascading Bistable Switches (CBS) - have been proposed. These models represent different mathematical realizations of the same core EMT circuit – miR-34/SNAIL/miR-200/ZEB1 – and both highlight that EMT is not an 'all-or-none' response (53,61–63), reminiscent of other similar examples of cellular plasticity (53) in tumor progression.

Here, we discuss the similarities and differences between these two models and present a new set of experiments that appear to align better with the TCS model. First, we show that the CD44s/ZEB1 feedback loop can underlie the self-activation of ZEB1. These results, coupled with the significantly different levels of ZEB1 and ESRP1 across E, E/M and M phenotypes in the NCI-60 cell line, and the single-cell co-expression of ZEB1 and E-cadherin in H1975 hybrid E/M cells, reinforce the role of ZEB1 in partial EMT. Second, we show that the TCS model can recapitulate the experimental phenomenon that the required concentration of exogenous TGF-β for the activation of SNAIL is lower than that for the activation of ZEB1, attributing to the stronger inhibition of ZEB1 by miR-200 than the inhibition of SNAIL by miR-34 (47,48). In addition, the TCS modeling results show that the tristability is more apparent for ZEB1 mRNA than that for ZEB1 protein, which again highlights the microRNA-mediated regulation. Third, the TCS modeling results suggest that the inhibition of miR-34 by ZEB1 can stabilize mesenchymal phenotype. Other interactions, although not exclusively, that can also help maintain mesenchymal phenotype is ZEB1 self-activation (**Figure S11**) and the autocrine miR-200/TGF-β loop, as also pointed out by Zhang et al. (32).

Further experiments that support the TCS model over the CBS model are related to the knockdown of FOXC2 – a transcription factor that is upregulated by SNAIL and directly activates the transcription of ZEB1 (36,49–51). Our experimental results show that knockdown of FOXC2 inhibits the gene expression changes concomitant with EMT, including ZEB1 activation, but does not affect SNAIL levels. According to the CBS model, FOXC2-KD cells can still undergo a partial EMT because upregulation of SNAIL levels is independent of FOXC2 and (high SNAIL, low ZEB1) levels are associated with a hybrid E/M state. But the TCS model proposes that knockdown of FOXC2 aborts EMT completely in the absence of ZEB1 (and FOXC2) expression irrespective of the high SNAIL levels, which is consistent with the experimental test. These results are reminiscent of observations in mouse mammary gland cells that decreased expression of ZEB1/2 is sufficient to re-establish the epithelial features (64) while knockdown of SNAIL is not (65), which implies an essential role for ZEB but not necessarily SNAIL in maintaining EMT in certain contexts. Further work on the interplay between ZEB1 and FOXC2 will give new insights into the role of miR-200/ZEB1 circuit during EMT.

The similarities and differences between the two models vis-à-vis the available experimental data calls for further quantitative analysis of EMT regulation in multiple contexts. The experimental data presented here favor the TCS model, but contextual differences in regulating EMT (2) might enable conditions where both models can reconcile different experimental observations.

Comparative analysis of these two mechanism-based models suggests several intriguing testable predictions that can help understand this multi-layered regulation of EMT/MET. First, ZEB1/2 mRNA, among other EMT regulators, should be measured at a single-cell level instead of population-averaged level at varying levels of an EMT-inducing signal such as TGF-β. The reason for this suggestion is three-fold: (a) the cell-to-cell variability tends to attenuate the observation of the intermediate level of ZEB1/2 mRNA (**Figure 3**), (b) the distribution of ZEB1/2 mRNA levels, as compared with ZEB1/2 protein levels, appears closer to being trimodal, and (c) although the cell lines belonging to different cancer types (29,44,66) have been classified into E, E/M, and M based on such ensemble measurements, recent studies have shown that cell lines can harbor phenotypic heterogeneity, therefore underlining the need to conduct single-cell studies (67). Second, in addition to these dose-response experiments, time-course measurements of levels of ZEB1/2, SNAIL1/2, miR-200, miR-34 and FOXC2, need to be performed to elucidate temporal dynamics of EMT.

The hybrid E/M state (partial EMT) has gradually drawn attention due to its proposed crucial role in tumor progression and organ fibrosis (11,68,69). The hybrid E/M state allows collective migration of CTCs as a cluster and these CTC clusters can evade immune attack (70) and often have much higher metastatic potential than the individually migrating CTCs (13). The TCS model can be utilized to identify certain 'phenotypic stability factors' (PSFs) (71), such as OVOL and GRHL2 (45), which can stabilize a 'metastable' partial EMT phenotype and thus being potential targets to 'break' the clusters of CTCs – the primary 'bad agents' of metastasis (11,13). In future, the landscape approach (72,73) could be utilized to quantify the stability of the hybrid E/M state and transition processes during EMT and MET. These insights can move us a step closer to understanding and eventually quantitatively predicting the population heterogeneity in an isogenic population (74).

In addition, cells undergoing EMT have been shown to acquire tumor-initiating and/or drug-resistance properties (often together referred to as 'stemness' in the context of Cancer Stem Cells) and to induce cell cycle arrest (11,54,69,75,76). By coupling the TCS model of EMT with the stemness model – LIN28/let-7/OCT4 in our previous work, we showed that 'stemness window' can slide on the 'EMT axis', or in other words, EMT-stemness correspondence can be fine-tuned by multiple players such as miR-200 that inhibits LIN28, and by PSFs such as OVOL (53). We also showed that JAG1 – a potential intercellular PSF – can mediate drug resistance (12). Such a dynamic positioning of the 'stemness window' can reconcile many apparently contradictory experiments about the effect of EMT/MET on stemness (77) – (a) complete EMT increases stemness (54,56), (b) MET increases stemness (78) and (c) partial EMT possesses maximum stemness (14,15). Future work to integrate the TCS model with regulatory networks for different 'hallmarks of cancer'(79) should be done to extend our understanding of EMT mechanisms and anti-metastasis strategies.

**Conclusions**

Our integrative modeling and experimental analyses of EMT/MET core network – miR-34/SNAIL/miR-200/ZEB1 - help distinguish mechanisms underlying EMT tristability, propose further experiments to decode the EMT dynamics more explicitly, and serve as a platform to identify certain 'underlying basic principles' pertaining to different hallmarks of cancer.

**Methods**

1. The model formulation and analyses

The detailed introduction of both the ternary switch (TCS) model and the cascading bistable switches (CBS) model can be found in the supplemental information Section 1&2.

(1) Deterministic analysis

The bifurcation diagrams of the EMT circuits are calculated by the Matcont package in Matlab. The temporal dynamics of the EMT circuits are calculated by the ode15s solver in Matlab.

(2) Stochastic simulation

For the single-cell simulation, a Langevin equation is used to describe the dynamic behaviors of the EMT circuits with the Gaussian white noise and the Euler-Maruyama method is used to integrate the equation (80).

For population-averaged calculation, 100 sets of parameters for the EMT circuit– miR-34/SNAIL/miR-200/ZEB1 were generated by parameter randomization (each parameter except for the hill coefficients is increased or decreased randomly up to 5% of the original value) to mimic the cell-cell variability. The random sampling follows a normal distribution. As for the hill coefficient, it follows the normal distribution within $[n-1, n+1]$, where $n$ is the original value.

The effective landscape (81,82) for the stable state $X$ (E or E/M or M) is defined by $E(X) = -ln(P(X))$, where $P(X)$ represents the probability of the stable state X to be observed.

2. Analysis of gene expression data from NCI-60 cell lines

The expression levels of ESRP1, ZEB1, VIM, CDH1, OVOL2 were downloaded from *Cellminer* (83) and categorized into E, E/M, and M sets based on CDH1/VIM ratio (29). Student's t-test were used to test the significance of difference in the expression levels. The Pearson's correlation coefficient is calculated between the expression level of ESRP1 and ZEB1, VIM, CDH1, OVOL2 respectively.

3. Immunofluorescence staining of NSCLC cell lines

NSCLC cell lines from initial authenticated cell passages, free from mycoplasma, were grown in RPMI 1640 with 10% FBS and 1% penicillin/streptomycin cocktail. For immunofluorescence, cells were fixed in 4% paraformaldehyde, permeabilized in 0.2% Triton X-100, and then incubated overnight with anti-rabbit ZEB1 (1:250; Cat. No. HPA027524, Sigma Aldrich), anti-rabbit CDH1 (1:200; Cat. No. 3195, Cell Signaling Technology), anti-rabbit vimentin (1:200; Cat. No. 5741, Cell Signaling Technology), anti-mouse CDH1 (1:100; Cat. No. 610182, BD Transduction) and anti-mouse vimentin (1:50; Cat. No. ab8978, Abcam). The primary antibodies were then detected with Alexa conjugated secondary antibodies (Life technologies). Nuclei were visualized by co-staining with DAPI.

4. RT-PCR analysis of EMT markers in NSCLC cell lines.

Total RNA was isolated following manufacturer's instructions using RNAeasy kit (Qiagen). cDNA was prepared using iScript gDNA clear cDNA synthesis kit (Bio-Rad). A TaqMan PCR assay was performed with a 7500 Fast Real-Time PCR System using TaqMan PCR master mix, commercially available primers, and FAM™-labeled probes for CDH1, ZEB1, FOXC2 and SNAIL and VIC™-labeled probes for 18S, according to the manufacturer's instructions (Life Technologies). Each sample was run in triplicate. Ct values for each gene were calculated and normalized to Ct values for 18S (ΔCt). The ΔΔCt values were then calculated by normalization to the ΔCt value for control.

5. Western Blot analysis of EMT markers in NSCLC cell lines

Cells were lysed in RIPA lysis assay buffer (Pierce) supplemented with protease and phosphatase inhibitor. The samples were separated on a 4–15% SDS-polyacrylamide gel (Biorad). After transfer to PVDF membrane, probing was carried out with primary antibodies and subsequent secondary antibodies. Primary antibodies were purchased from the following commercial sources: anti-CDH1 (1:1000; Cell Signaling Technology), anti-vimentin (1:1000; Cell Signaling Technology), anti-ZEB1 (1:1000; Cell Signaling Technology), anti-SNAIL (1:1000; Cell Signaling Technology), anti-FOXC2 (1:2000; Bethyl Laboratories) and anti-GAPDH (1:10,000; Abcam). Membranes were exposed using the ECL method (GE Healthcare) according to the manufacturer's instructions.

6. Western blot analysis of EMT markers in HMLE cells

Authenticated immortalized human mammary epithelial (HMLE) cells expressing empty vector (pWZL), Snail, Snail shControl, or Snail shFOXC2 cells were cultured in MEGM media as previously described (51). For immunoblotting, cells were lysed in RIPA buffer, run on an 8% gel, and transferred onto nitrocellulose. Primary antibodies were incubated with the membrane overnight at 4°C and included β-actin (Santa Cruz Biotechnology, Dallas, TX, USA), FOXC2 (developed by Dr Naoyuki Miura, Hamamatsu University School of Medicine, Hamamatsu, Japan), E-cadherin (BD Biosciences, San Jose, CA, USA; 61081), Fibronectin (BD Biosciences; 610077), Vimentin (V9, Thermo Fisher, MA5-11883), and ZEB1 (Santa Cruz Biotechnology, Dallas, TX, USA).

**List of abbreviations**

Epithelial-to-Mesenchymal Transition (EMT)

Mesenchymal-to Epithelial-Transition (MET)

Ternary Chimera Switch (TCS)

Cascading Bistable Switches (CBS)

Circulating Tumor Cell (CTC)

Phenotypic Stability Factor (PSF)

**Competing interest**

The authors declare no conflict of interest.

**Funding**

This work was supported by the National Science Foundation (NSF) (DMS-1361411 and Grants PHY-1427654) and by the Cancer Prevention and Research Institute of Texas (CPRIT) (Grants R1110 and R1111). M.L. has a training fellowship from the Keck Center for Interdisciplinary Bioscience Training of the Gulf Coast Consortia (CPRIT Grant RP140113).

**Authors' contributions**

Conceptualization, H.L., J.O., M.L., M.K.J., and E.B.-J; Methodology, D.J., M.K.J., S.C.T., P.D.H., B.H., M.L., M.C. and E.R.-P.; Investigation, D.J., M.K.J., S.C.T., P.D.H., B.H., M.L., M.C., E.R.-P., E.B.-J, J.O., S.M.H., S.A.M. and H.L.; Writing, D.J., M.K.J., S.C.T., P.D.H., B.H., M.L., E.B.-J, J.O., S.M.H., S.A.M. and H.L.; Funding Acquisition, S.M.H., S.A.M., J.O. and H.L.; Supervision, S.M.H., S.A.M., J.O. and H.L..

# Figure legends

**Figure 1. Schematic representations of the ternary chimera switch (TCS) and the cascading bistable switches (CBS) models.** The core EMT regulatory network in the TCS (**A**) model and the CBS (**B**) model. Bifurcation diagrams of mesenchymal (M) marker levels in response to exogenous TGF-β in the TCS model (**C**) and the CBS model (**D**). Bifurcation diagrams of ZEB1 mRNA levels in response to SNAIL for the miR-200/ZEB1 circuit in the TCS model (**E**) and the CBS model (**F**). In (A) and (B), $\mu_{34}$ represents miR-34 and $\mu_{200}$ represents miR-200. Solid arrows represent transcriptional activations and solid bar-headed arrows represent transcriptional inhibitions. Dashed bar-headed arrows represent microRNA-mediated regulations. The circled arrow along with ZEB1 in (A) represent ZEB1 self-activation and the circled bar-headed arrow along with SNAIL in (A) and (B) represent SNAIL self-inhibition. In (C) and (D), blue solid lines represent stable states and red dotted lines represent unstable states. The dotted arrow represents transitions between different stable states. The cell phenotype corresponding to each stable state is labeled and also shown in cartoon.

**Figure 2. The CD44s/ZEB1 feedback loop enables the miR-200/ZEB1 circuit to function as a three-way switch.** (**A**) Top panel: The SNAIL-driven miR-200/ZEB1 circuit including the CD44s/ZEB1 feedback loop. Bottom panel: Bifurcation diagram of ZEB1 mRNA levels for the miR-200/ZEB circuit in response to SNAIL. (**B**) Top panel: The $S_1$-$S_2$-driven miR-200/ZEB1 circuit including the CD44s/ZEB1 feedback loop. $S_1$ represents a transcriptional inhibition signal on miR-200 and $S_2$ represents a transcriptional activation signal on ZEB1. Bottom panel: The phase diagram (a two-parameter bifurcation) of the miR-200/ZEB1 circuit driven by signals $S_1$ and $S_2$. In (A) and (B), the dotted bar-headed arrows represent the alternative splicing of CD44

mRNAs by ESRP1. Different colors in the bifurcation diagrams represent different co-existences of stable states. For example, the blue colored region marks the tristable phase – {E, E/M, M}, where all three phenotypes - E, E/M and M can be the stable states. In (B), the region marked by black dots in phase diagram represents the parameter region of $S_1$ and $S_2$ for the existence of hybrid E/M phenotype – existing either alone - {E/M} or in combination with other stable states – {E, E/M}, {M, E/M} and {E, E/M, M}. (**C**) Top panel: Relative gene expression levels of ZEB1 and ESRP1 in epithelial (n=11), hybrid E/M (n=11) and mesenchymal (n=37) cell lines from NCI-60. Bottom panel: Pearson's correlations between gene expression of ESRP1 and ZEB1, VIM, CDH1 and OVOL2. '*' represents $P$ value $\leq 0.05$. '**' represents $P$ value $< 0.05$. '***' represents $P$ value $< 0.0001$. (**D**) Immunofluorescence images showing different expression patterns of EMT markers in NSCLC cell lines. In the first column, blue is for DAPI, red is for ZEB1 and green is for CDH1. In the second column, blue is for DAPI, red is for CDH1 and green is for VIM. (**E**) mRNA levels of CDH1, VIM, SNAIL and ZEB1 in NSCLC cell lines. (**F**) Protein levels of CDH1, VIM, SNAIL, ZEB1 and FOXC2 in NSCLC cell lines. In (E) and (F), H820 and H1437 are epithelial cell lines, H1299 and H2030 are mesenchymal cell lines, H1975 is hybrid E/M cell line and H1944 is a mixture of E and M cells.

**Figure 3. Simulation of the responses of SNAIL and ZEB1 to EMT-inducing signal $S_A$ by the TCS model on single-cell and population levels.** For each figure, six different levels of $S_A$ are chosen - $S_A$ = 0, 20, 36, 49, 64, 90 ($*10^3$ molecules). The distributions of mRNA and protein levels of SNAIL and ZEB in response to different levels of $S_A$ are calculated by Langevin simulation. For (**B**) and (**C**), white noise to mimic the fluctuations inside one cell is considered. For (**E**) and (**F**), parameter randomization to mimic the cell-cell variability is included. The trimodal distribution of ZEB1 mRNA levels when $S_A = 49*10^3$ molecules during EMT is highlighted in both (**C**) and (**F**).

**Figure 4. The inhibition of miR-34 by ZEB1 promotes the 'irreversibility' of complete EMT, as elucidated by the TCS model.** (**A**) Bifurcation diagram of ZEB1 mRNA levels in response to activation signal on SNAIL ($S_A$) when $\lambda_{Z,\mu_{34}}$ = 0, 0.2, 0.4, 0.6, 0.8, 1. $\lambda_{Z,\mu_{34}}$ is the fold-change parameter which represents the strength of ZEB1 inhibition on miR-34. $\lambda_{Z,\mu_{34}} = 1$ represents that there is no inhibition from ZEB1 to miR-34. $\lambda_{Z,\mu_{34}} = 0$ represents strong inhibition from ZEB1 to miR-34. The dotted line in the bifurcation highlights the value of activation signal $S_A = 50 * 10^3$ molecules, which is used to calculate the effective landscape $(-log(P))$ as shown in (**B**). (**B**) Effective landscape of EMT for $\lambda_{Z,\mu_{34}} = 1$ (no feedback), $\lambda_{Z,\mu_{34}} = 0.6$ (Intermediate feedback) and $\lambda_{Z,\mu_{34}} = 0$ (strong feedback). The stable state corresponding to each basin is labeled. The larger the blue area surrounding a particular state is, the more frequently that stable state can be achieved.

**Figure 5. The autocrine miR-200/TGF-β loop can maintain the mesenchymal phenotype.** (**A**) The framework of TCS model with the autocrine miR-200/TGF-β loop. (**B**) Phase diagram of the circuit shown in (A) in response to an activation signal on SNAIL ($S_A$) and the production rate of endogenous TGF-β ($g_{m_{TGF-\beta}}$). Two different values of $g_{m_{TGF-\beta}}$ are highlighted, as they are the values used to calculate the bifurcation in (**C**) and (**D**). (**C**) Bifurcation of ZEB1 mRNA levels in response to $S_A$ when $g_{m_{TGF-\beta}} = 2$ molecules/hour. (**D**) Bifurcation of ZEB1 mRNA level in response to $S_A$ when $g_{m_{TGF-\beta}} = 10$ molecules/hour. The arrow in (D) highlights that the mesenchymal state can be maintained without EMT-inducing signal $S_A$ when the basal

production rate of endogenous TGF-β is high. Different colors in (**B-D**) represent different combinations of possible stable states.

**Figure 6. Overexpression of SNAIL is not sufficient to initiate EMT in absence of ZEB1 and FOXC2.** (**A**) Integrating FOXC2 to the TCS framework – TCS-FOXC2. (**B**) Top panel: Bifurcation diagram of the TCS-FOXC2 circuit when FOXC2 is present, as reflected by $S_I = 0$. Bottom panel: Bifurcation diagram of the TCS-FOXC2 circuit when FOXC2 is absent, as reflected by $S_I = 2 * 10^5$ molecules. (**C**) Western blot analysis of EMT marker expression upon SNAIL overexpression in HMLE cells, and FOXC2 suppression in HMLE–Snail cells respectively. (**D**) The proposed EMT regulatory framework integrating transcriptional regulation by FOXC2.

# Figure 1

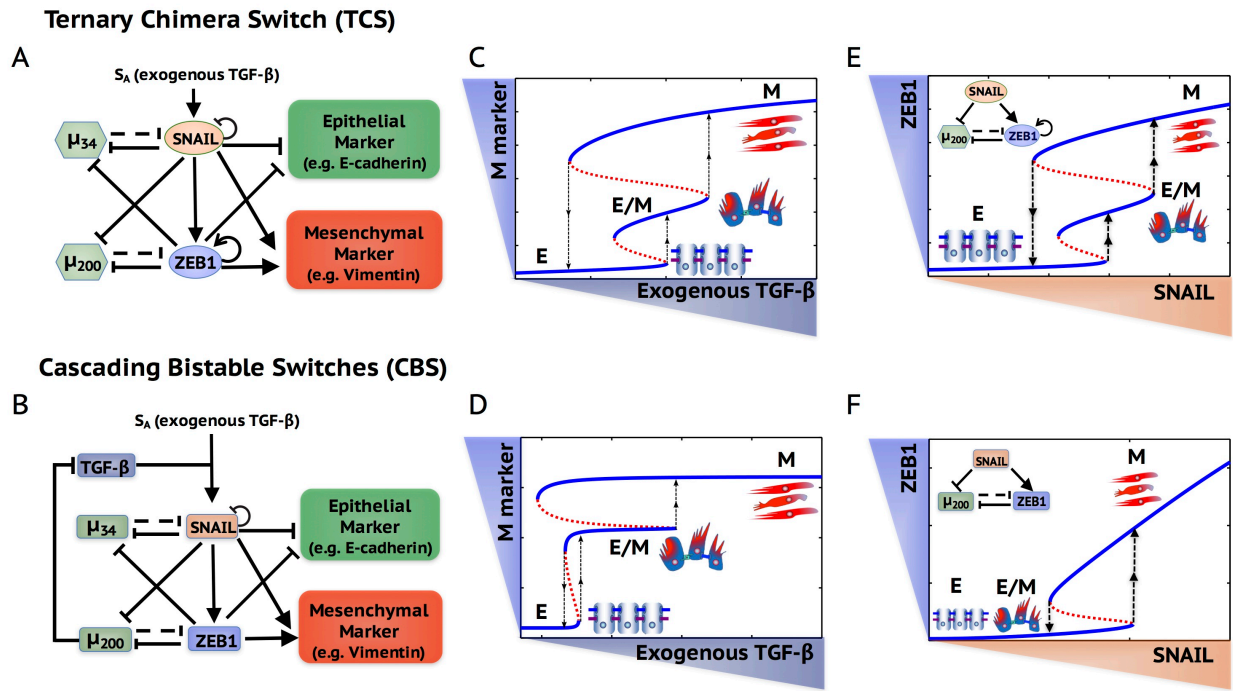

**Figure 2**

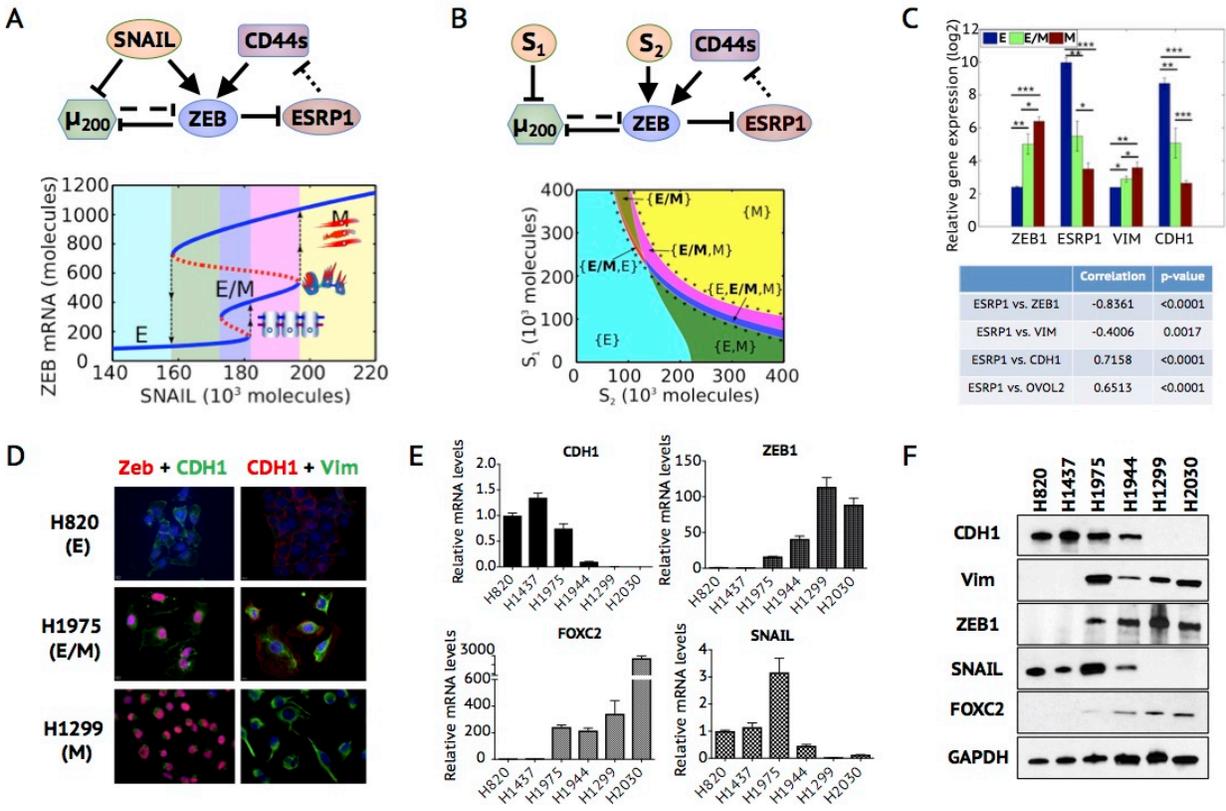

**Figure 3**

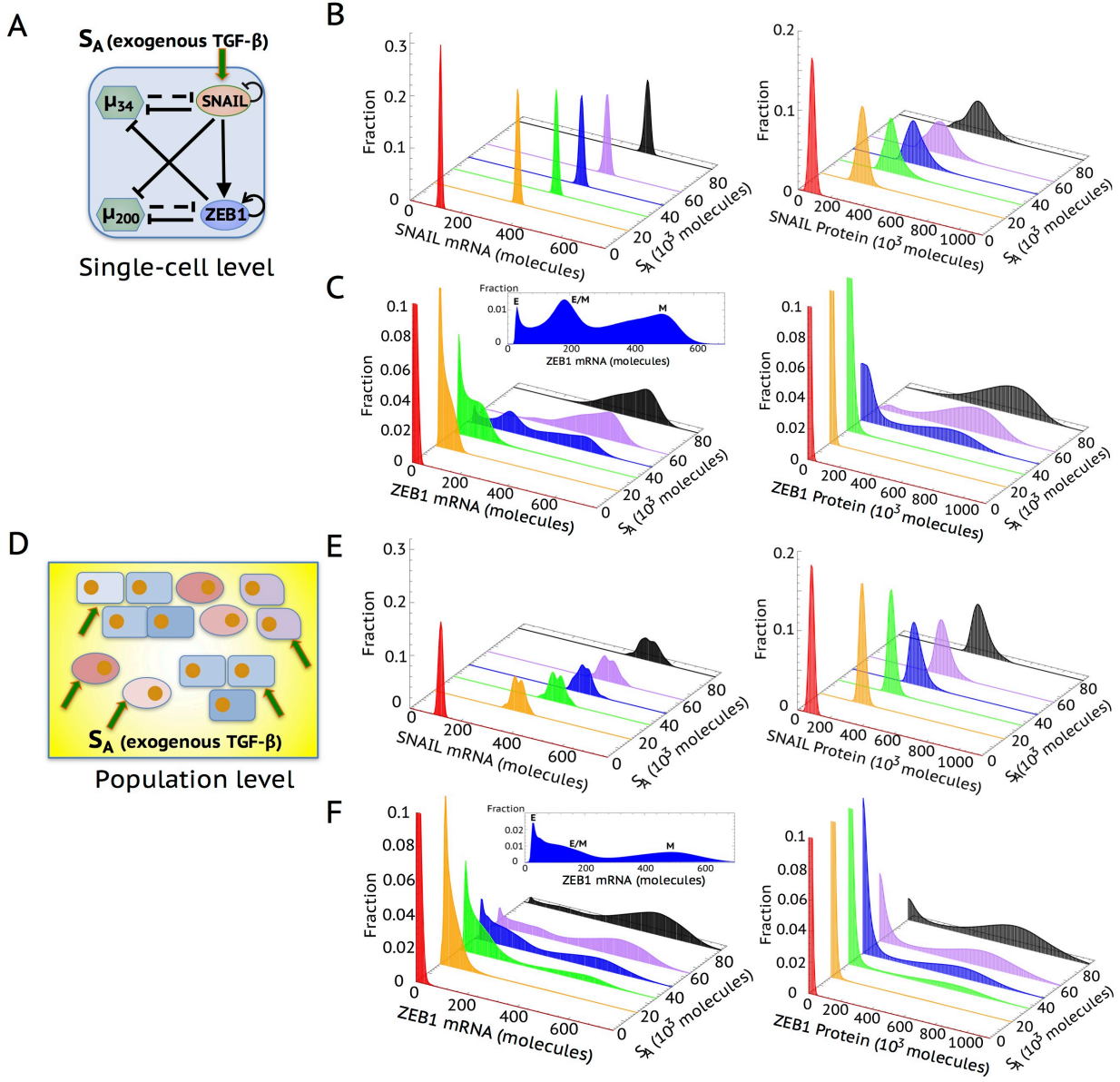

**Figure 4**

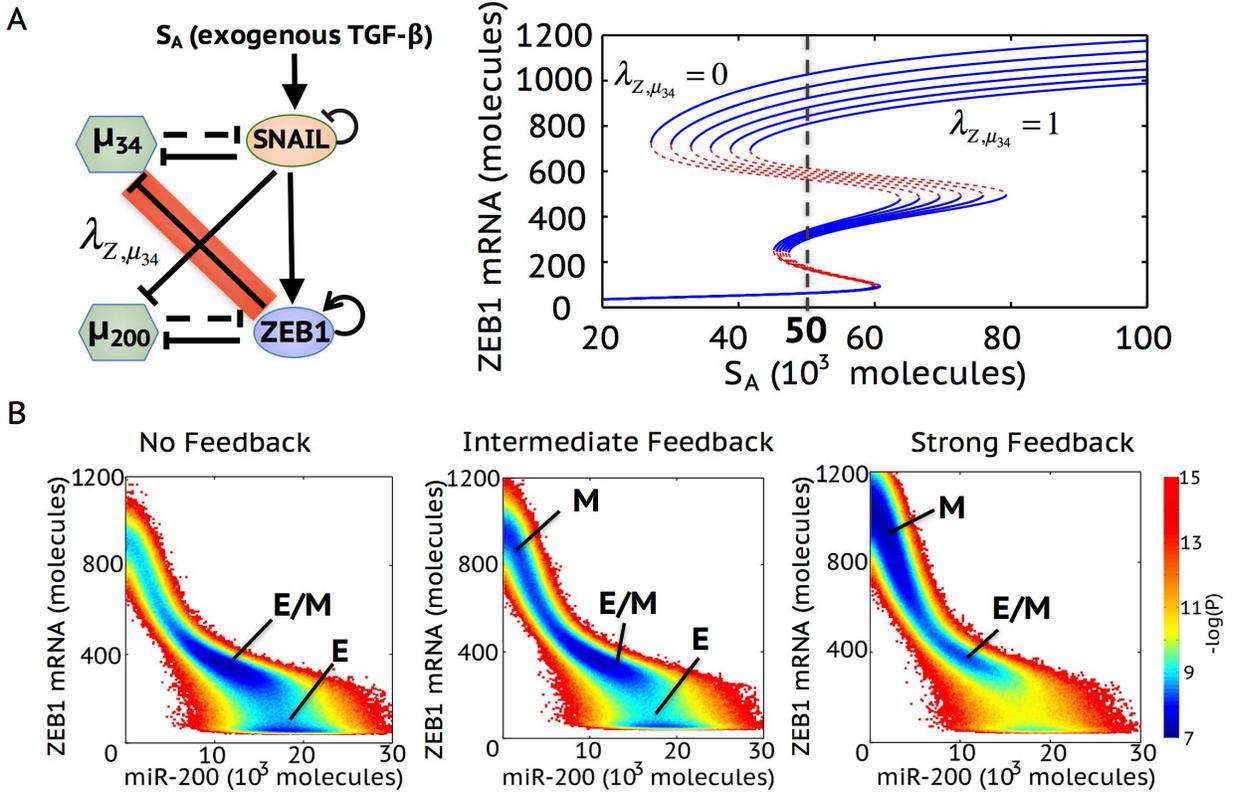

**Figure 5**

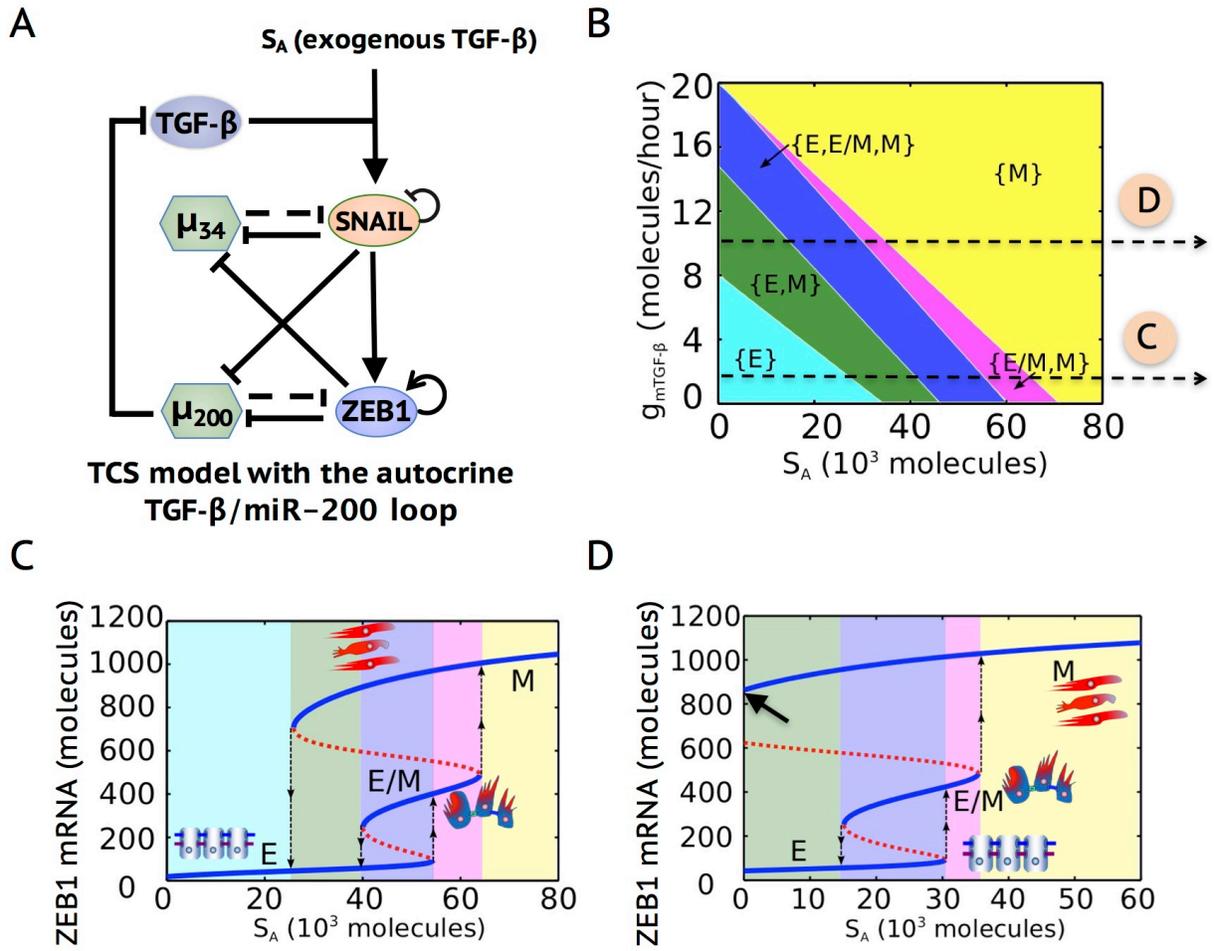

**Figure 6**

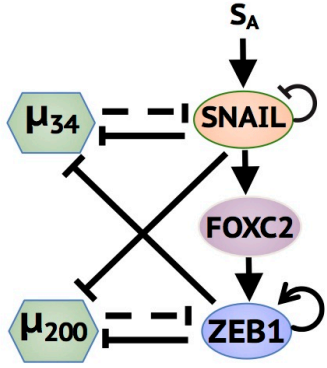
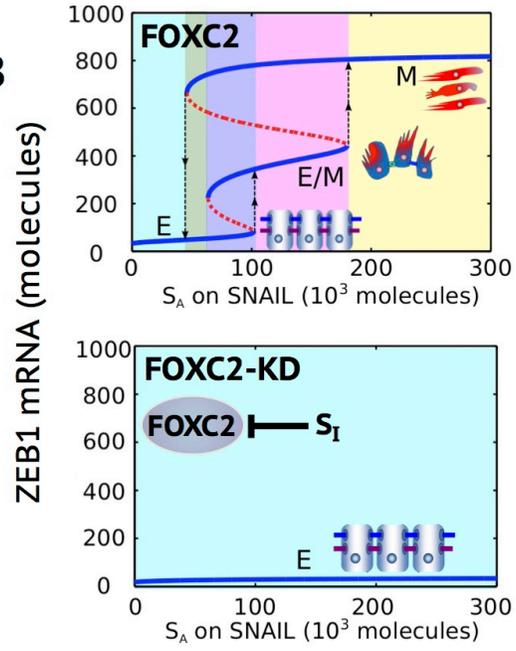
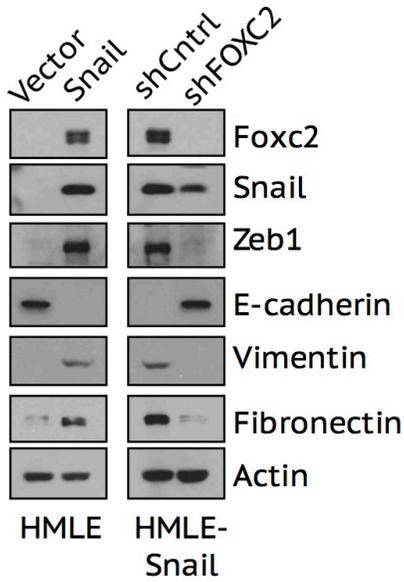
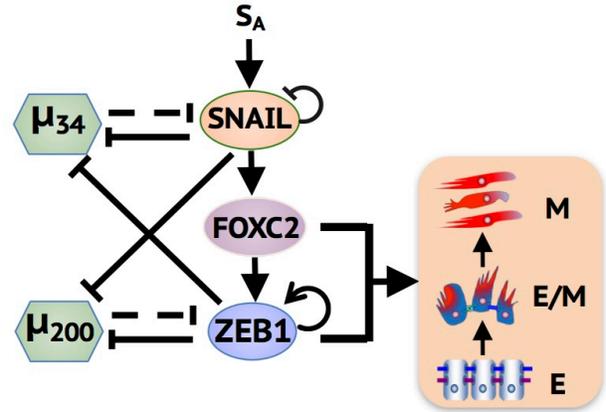